\documentclass[sigconf]{acmart}

\renewcommand\footnotetextcopyrightpermission[1]{} 
\AtBeginDocument{%
  }

\acmConference[GenAIRecP@KDD'25]{%
  Proceedings of the 31st ACM SIGKDD Conference on Knowledge Discovery and Data Mining%
}{August 4, 2025}{Toronto, Canada}

\acmBooktitle{%
  Second Workshop on Generative AI for Recommender Systems and Personalization at the ACM Conference on Knowledge Discovery and Data Mining (GenAIRecP@KDD 2025),
August 4, 2025, Toronto, Canada%
}

\usepackage{svg}
\usepackage{graphicx}
\usepackage{epstopdf}
\begin{document}

\title{End-to-End Personalization: Unifying Recommender Systems with Large Language Models }

\author{Danial Ebrat}
\email{Ebrat@uwindsor.ca}
\affiliation{%
  \institution{University of Windsor}%
  \city{Windsor}%
  \state{Ontario}%
  \country{Canada}%
} %
\author{Tina Aminian}
\email{aminiant@uwindsor.ca}
\affiliation{%
  \institution{University of Windsor}%
  \city{Windsor}%
  \state{Ontario}%
  \country{Canada}%
}%

\author{Sepideh Ahmadian}
\email{ahmadia3@uwindsor.ca}
\affiliation{%
  \institution{University of Windsor}%
  \city{Windsor}%
  \state{Ontario}%
  \country{Canada}%
} %

\author{Luis Rueda}
\email{lrueda@uwindsor.ca}
\affiliation{%
  \institution{University of Windsor}%
  \city{Windsor}%
  \state{Ontario}%
  \country{Canada}%
} %
\renewcommand{\shortauthors}{Ebrat et al.}








\begin{abstract}
  Recommender systems are essential for guiding users through the vast and diverse landscape of digital content by delivering personalized and relevant suggestions. However, improving both personalization and interpretability remains a challenge, particularly in scenarios involving limited user feedback or heterogeneous item attributes. In this article, we propose a novel hybrid recommendation framework that combines Graph Attention Networks (GATs) with Large Language Models (LLMs) to address these limitations. LLMs are first used to enrich user and item representations by generating semantically meaningful profiles based on metadata such as titles, genres, and overviews. These enriched embeddings serve as initial node features in a user–movie bipartite graph, which is processed using a GAT-based collaborative filtering model. To enhance ranking accuracy, we introduce a hybrid loss function that combines Bayesian Personalized Ranking (BPR), cosine similarity, and robust negative sampling. Post-processing involves reranking the GAT-generated recommendations using the LLM, which also generates natural-language justifications to improve transparency. We evaluate our model on benchmark datasets, including MovieLens 100k and 1M, where it consistently outperforms strong baselines. Ablation studies confirm that LLM-based embeddings and the cosine similarity term significantly contribute to performance gains. This work demonstrates the potential of integrating LLMs to improve both the accuracy and interpretability of recommender systems. 
\end{abstract}

\maketitle

\section{Introduction}
The growing demand for personalized, context-aware, and interpretable recommendation systems has led to a surge of interest in integrating LLMs into the recommendation pipeline. Conventional recommender systems struggle with limitations such as data sparsity, shallow contextual understanding, and the need for extensive manual feature engineering. LLMs represent a paradigm shift, offering richer feature representations, adaptive reasoning, and the ability to enhance transparency throughout the recommendation process. 

Recent research has explored various avenues through which LLMs contribute to recommendation systems, including data augmentation, feature generation, re-ranking, and explanation generation. For instance, a comprehensive taxonomy of LLM-augmented recommenders, highlighting their role across different stages of the pipeline, provided by \cite{DBLP:journals/tkde/ZhaoFLLMWWWZTL24}. In addition, the authors of \cite{DBLP:journals/www/WuZQWGSQZZLXC24,DBLP:journals/corr/abs-2306-05817}, emphasize the integration of LLMs in machine learning workflows, underscoring their utility in preprocessing, knowledge alignment, and ranking optimization. 

Building upon this body of work, we propose a novel recommendation framework that strategically incorporates LLMs into both preprocessing and post-processing stages. In the preprocessing phase, semantically enriched user and item profiles are generated through multi-turn interactions and schema-aligned transformations. Post-processing involves an LLM-driven re-ranking module that re-evaluates top-N recommendations for semantic alignment and diversity. At the core of the pipeline, a GAT is employed to capture user-item interactions using LLM-derived embeddings, thereby enhancing cold-start robustness and contextual sensitivity. 

This section reviews relevant literature across three key dimensions: (1) the use of LLMs for feature engineering and representation learning, (2) LLMs for ranking refinement and interpretability, and (3) the use of Graph Attention Networks in recommender systems. 

\subsection{LLMs for Feature Engineering and Representation Learning }
Feature engineering is central to the performance of recommender systems, traditionally requiring substantial domain knowledge and manual effort. The advent of LLMs has shifted this paradigm by enabling automated, context-aware transformation of raw textual data into semantically enriched representations. These capabilities are particularly valuable in scenarios with sparse or noisy inputs, where traditional techniques struggle to capture nuanced relationships. 

Recent research has explored a variety of LLM-based approaches for representation enhancement. KAR proposes auxiliary feature generation for user-item modeling, though its reliance on static feature construction reduces adaptability \cite{DBLP:conf/recsys/XiLLCZZCT0024}. SAGCN employs chain-based prompting to reveal semantic relationships, although its performance is highly sensitive to prompt design and consistency 
\cite{DBLP:journals/corr/abs-2312-16275}. CUP addresses input length limitations through compact summarization, yet often sacrifices fine-grained user preferences \cite{DBLP:journals/corr/abs-2311-01314}. In domain-specific applications, LLaMA-E and EcomGPT apply LLMs for attribute extraction, achieving promising results within narrow verticals, yet their generalizability remains limited \cite{DBLP:journals/corr/abs-2308-04913,DBLP:conf/aaai/LiMWHJ0X0J24}. Additionally, LLMs have been applied in preprocessing pipelines for knowledge graph completion \cite{DBLP:journals/corr/abs-2305-09858, DBLP:journals/corr/abs-2401-08217, DBLP:journals/corr/abs-2308-10835, DBLP:conf/wsdm/WeiRTWSCWYH24}, text refinement \cite{DBLP:conf/aaai/DuL0W0Z0Z24, DBLP:conf/www/LiuCZDWLL0024, DBLP:journals/corr/abs-2307-02157}, and synthetic data generation \cite{DBLP:conf/emnlp/LiZL023}. While these methods mitigate data sparsity and cold-start issues, they risk introducing semantic noise or bias if outputs are not rigorously validated. 

In this paper, we introduce End-to-End Personalization, a unified pipeline that integrates GATs and LLMs to enhance recommendation quality. Our methodology employs LLMs in a principled, schema-aligned manner to ensure semantic coherence, consistency, and interpretability throughout the recommendation process. This structured preprocessing pipeline enhances embedding initialization for graph-based models, offering improvements in both personalization depth and generalizability. Building on our prior framework \cite{DBLP:journals/corr/abs-2405-13362}, which demonstrated the effectiveness of LLMs in generating dynamic, explainable feedback for evolving user states, we further refine their role in pretraining representations for robust recommendation.

\subsection{LLMs for Ranking and Interpretability}
Ranking plays a critical role in shaping user experience, influencing not only which items are recommended but also the order in which they are prioritized. Conventional models, including matrix factorization \cite{DBLP:conf/www/HeLZNHC17}, sequence-based predictors \cite{DBLP:journals/fcsc/ChengLZLZC24,DBLP:conf/recsys/Liu0LTW23}, and GNNs \cite{DBLP:conf/sigir/Wang0WFC19}, offer strong performance while often lacking transparency and interpretability—qualities increasingly demanded in sensitive domains such as health, finance, and education.

Existing LLM-based ranking systems fall into two primary categories. Scoring models, such as E4SRec and ClickPrompt use modified architectures to output relevance scores, offering efficiency while limiting the expressive capabilities of the underlying model \cite{DBLP:journals/corr/abs-2312-02443,DBLP:conf/www/LinCWXQDZTY024}. Two-tower frameworks like CoWPiRec are scalable, though model shallow interactions, relying on fixed similarity metrics \cite{DBLP:conf/icdm/00040LXM0ZZFD23}. Classification approaches like TALLRec reformulate ranking as a prediction task, yet often struggle with score calibration in multi-item settings \cite{DBLP:conf/recsys/BaoZZWF023}. Generative models, such as LANCER and LlamaRec, offer more flexibility, yet are prone to hallucination and are constrained by retrieval quality \cite{DBLP:conf/dasfaa/JiangQCLLZZZLLG24, DBLP:journals/corr/abs-2311-02089}. 

To address these limitations, our approach combines graph-based representation learning with a lightweight LLM reranker. By decoupling ranking from generation and leveraging semantically structured profiles, we achieve higher interpretability, lower computational cost, and stronger alignment with user intent. 
\subsection{Graph Attention Networks in Recommendation Systems }
Graph-based collaborative filtering methods have become foundational in modern recommender systems due to their ability to model high-order interactions in user–item bipartite graphs. Traditional techniques like matrix factorization (e.g., SVD, ALS) struggle to incorporate contextual signals and often underperform in cold-start scenarios \cite{golub1968least,DBLP:journals/simods/LeeS23}. Early propagation-based models, including ItemRank and BiRank, improved upon this by diffusing preferences across the graph structure while lacked trainable parameters, reducing their expressiveness \cite{DBLP:conf/ijcai/GoriP07, DBLP:journals/tkde/HeGKW17}. 

The introduction of Graph Neural Networks (GNNs) transformed this landscape by enabling message-passing architectures to capture both local and global interaction patterns. GC-MC, PinSage, and SpectralCF demonstrated how incorporating node features and graph topology can enhance recommendation accuracy \cite{DBLP:conf/bibm/YangZDXDLSJZL23,DBLP:conf/kdd/YingHCEHL18,DBLP:conf/recsys/ZhengLJZY18}. NGCF introduced explicit multi-hop connectivity through stacked convolutions, though its complexity raised concerns about overfitting \cite{DBLP:conf/www/HeLZNHC17}. LightGCN addressed this by simplifying the architecture, removing activation functions, and emphasizing pure neighborhood aggregation \cite{DBLP:conf/sigir/0001DWLZ020}. GATs further extend this paradigm by assigning adaptive weights to neighbors during  message passing, allowing for more selective and context-sensitive representation learning. 

 Attention-based recommendation models like IGAT and TKGAT improved flexibility at the cost of scalability and interoperability \cite{DBLP:journals/eswa/ElahiAARNHW25, DBLP:conf/adma/ZhangLWCG23}. In this work, we adopt a lightweight GAT architecture that integrates LLM-derived semantic profiles as node features, combining the relational strength of GNNs with the contextual richness of LLMs. This fusion improves performance in sparse regimes and supports personalization through adaptive, meaningful attention. performance in sparse regimes and supports personalization through adaptive, meaningful attention. 

\begin{figure*}[ht]
  \centering
  \includegraphics[width=0.60\textwidth]{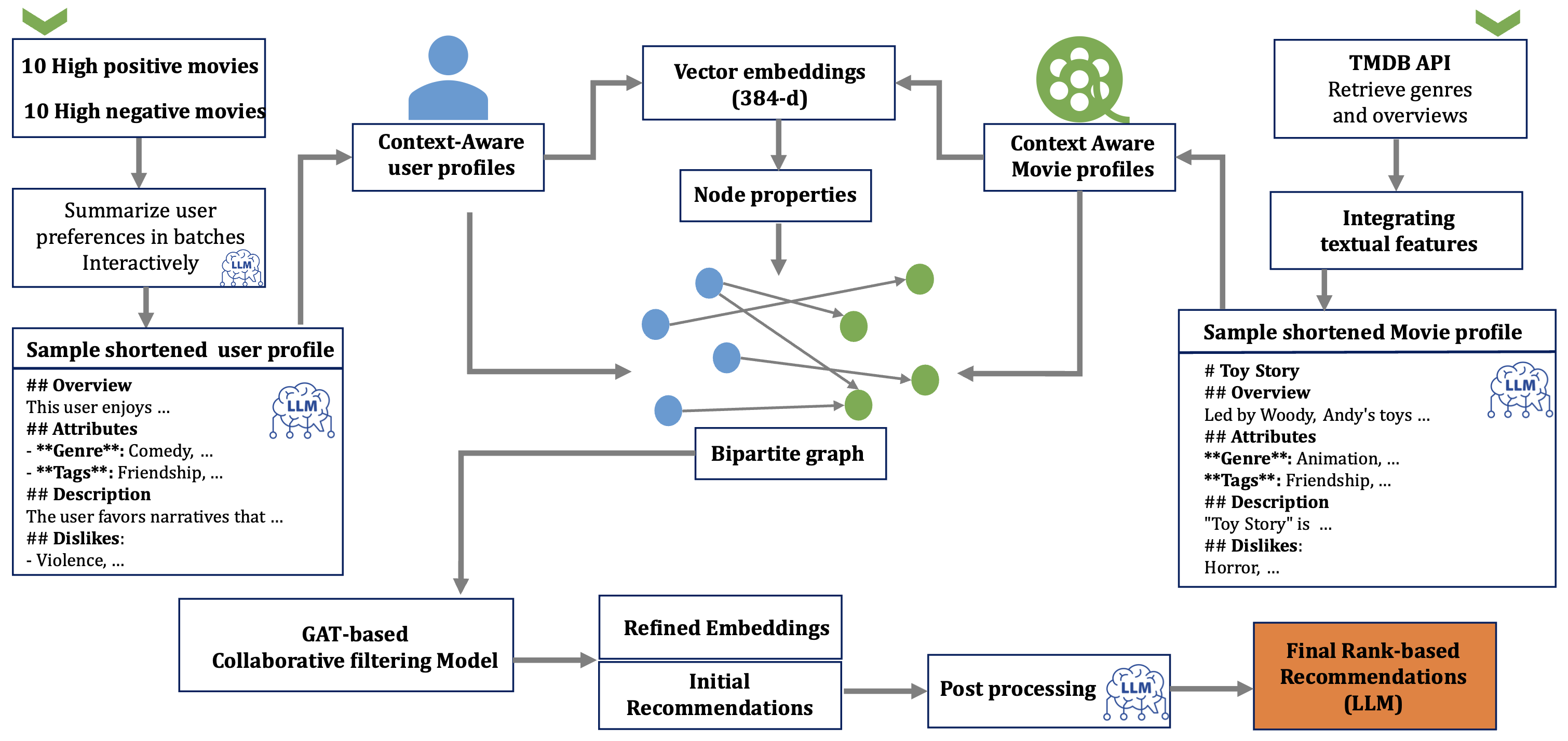}
  \caption{Schematic view of the pipeline utilized by the proposed method.}
  \Description{Enjoying the baseball game from the third-base
  seats. Ichiro Suzuki preparing to bat.}
  \label{fig:teaser}
\end{figure*}
 
 \subsection{Contributions}
 This paper introduces a unified architecture that integrates LLMs into both the preprocessing, which also directly affects model training, and post-processing stages of the recommendation pipeline. 
\begin{itemize}
    \item \textbf{Semantic Profiling via LLMs}: We propose a structured preprocessing pipeline that transforms raw movie and user data into semantically enriched profiles using multi-turn interactions and schema alignment. 

    \item \textbf{Iterative User Preference Modeling}: A novel multi-turn LLM dialogue system is introduced to incrementally capture complex user preferences, overcoming token constraints while maintaining continuity and coherence.

    \item \textbf{Structured Embedding Initialization for Graph Models}: The enriched profiles are embedded and used to initialize a GAT, including a combined loss function that integrates BPR with a cosine similarity term and robust negative sampling to optimize ranking performance and the alignment of semantically similar embeddings, enhancing representation quality in sparse data conditions and cold-start scenarios. 

    \item \textbf{Post-Hoc Reranking and Explainability}: An LLM-based reranker evaluates top-N recommendations, offering fine-grained diversity and semantic alignment alongside human-interpretable explanations. 
\end{itemize}
Collectively, these components yield a transparent, generalizable, and high-performing recommendation framework that addresses longstanding challenges in feature sparsity, cold-start handling, and interpretability. By bridging LLM-driven reasoning with graph-based relational learning, this work sets a new benchmark for scalable and explainable recommender systems. 


\section{Methodology}
This section outlines our proposed methodology, systematically structured into three main distinct yet interconnected phases: Preprocessing and LLM-Based Profile Generation, Collaborative Filtering Model and training procedures, and LLM-Based Post-processing, and Explainability. Each phase emphasizes clear integration and justification of LLMs within our recommender system pipeline.   Figure 1, depicts a schematic view of the methodology pipeline and how we integrate these steps.  

\subsection{Preprocessing and Profile Generation }

We utilized the widely recognized MovieLens 100K and 1M datasets, enhancing them through additional metadata acquired from The Movie Database (TMDB) API. Specifically, movie titles, genres, and textual overviews were integrated to form coherent, semantically enriched descriptions. This initial preprocessing stage provides the foundational textual resources necessary for subsequent advanced semantic analysis. Furthermore, standardized normalization and cleaning techniques were applied to ensure data consistency and quality. 
\subsubsection{Item Profile Generation} 
Movie profiles were enriched through an advanced LLM-driven process. The textual metadata (titles, genres, overviews) served as inputs to an LLM agent tasked with extracting structured, nuanced descriptors. These descriptors included specific narrative elements (e.g., "time travel," "heist scenarios") and character-driven attributes (e.g., "anti-hero," "strong female protagonist"), significantly enhancing profile specificity and semantic depth.
\subsubsection{User Profile Generation}
To capture complex and multifaceted user preferences accurately, an iterative, conversational refinement process was developed leveraging multi-turn LLM interactions. We begin by selecting the user's 10 highest-rated (4-5) and 10 lowest-rated (1-2) movies, processed incrementally in batches of 5. Each batch triggered sequential LLM interactions, with prompt engineering explicitly incorporating previously refined profile elements to ensure continuity and semantic consistency across conversational turns. This iterative methodology effectively addressed context-length constraints, resulting in highly nuanced user profiles reflective of sophisticated preference patterns.
\subsubsection{Structured Schema Alignment}
Both movie and user profiles followed an aligned, structured schema emphasizing semantic coherence. This schema included consistent and clearly defined, enabling accurate comparison, vector embedding representation, and robust matching accuracy between user interests and movie characteristics.  

The United profile structure using Markdown language contains: 
\begin{itemize}
    \item \textbf{Overview}: Summary capturing the core narrative or thematic essence.
    \item \textbf{Attributes}: Concise genres and descriptive tags relevant to content. 
    \item \textbf{Description}: Expanded narrative and character-focused insights. 
    \item \textbf{Dislikes}: Explicitly identified non-relevant attributes. 
\end{itemize}

\subsection{Model Embedding Initialization}
Semantic vectors for the textual profiles were first generated with the pretrained SentenceTransformer allMiniLML6v2 \cite{DBLP:conf/nips/WangW0B0020}, producing 384dimensional embedding representations that seeded a bipartite useritem graph. We then refine these embeddings with a threelayer Graph Attention Network tailored for collaborative filtering. Each layer featured 64 hidden units, four attention heads, layer normalization, LeakyReLU activations, residual skips, and dropout, enabling bidirectional message passing so that user and item nodes updated one another simultaneously. Edges captured explicit feedback—ratings ~$\geq$~ 4 as positive and ~$\leq$~ 2 as negative—while neutral scores were omitted to limit noise. Training optimized a blend of  BPR loss and a cosinebased alignment regularizer that pulls embeddings of positively rated useritem pairs closer together. Using AdamW with adaptive learning rates, weight decay, and early stopping secured stable convergence, and final relevance scores were computed via the dot product of the refined user and item vectors. 

\subsection{Post-processing and Explainability }
 Following the GAT-based collaborative filtering stage, we introduce a series of LLM-driven post-processing methods designed to refine recommendation rankings and generate transparent, user-facing rationales. Each method begins with the initial candidate pool produced by graph-based signals, subsequently refined through semantic reasoning by an LLM. The refined rankings are then fused with the original GAT scores using an 80:20 weighted hybrid scheme—assigning 80\% weight to the LLM output—ensuring enhanced semantic alignment without disregarding interaction patterns. 

   We implemented and evaluated several post-processing variants to rerank the top 20 recommendations for each user: 
\begin{itemize}
    \item \textbf{Prompt-level Re-ranking}: A context-rich prompt containing the user's genre preferences, thematic leanings, recent interactions, and detailed metadata of candidate films is provided to the LLM. The model performs a single-pass semantic alignment and re-ranking of all 20 candidates simultaneously. 
    \item \textbf{Pairwise BST-based Re-ranking}: Leveraging a balanced Binary Search Tree (BST), the LLM conducts pairwise comparisons, evaluating two items at a time against user preferences to determine relative suitability. This method significantly reduces redundant comparisons by preserving logical order.
    \item \textbf{Batch-of-5 Re-ranking with Overlaps} items from bigger pool are partitioned into overlapping batches of five to maintain prompt conciseness and minimize hallucinations. Each batch is independently re-ranked by the LLM, and global rankings are subsequently merged based on overlapping results to ensure consistency and accuracy. 
    \item \textbf{Relevancy Scoring Across Batches}: Items are organized into batches of five from bigger pool, each batch evaluated separately by the LLM, which assigns relevancy scores (0-100) to individual movies. To mitigate scoring biases, each film appears in three different batches. Final scores are averaged across the batches. 
\end{itemize}

For each post-processing variant, the same prompt context is reused immediately after re-ranking to generate succinct, natural-language explanations articulating the rationale behind each recommendation. Explanations explicitly reference shared narrative themes, stylistic elements, and cast or crew relevance, enhancing transparency without additional inference overhead. 

We utilized both the OpenAI 4o-mini and the o4-mini models and the open-source Gemma-3 4B model for all methodologies. This comparative analysis facilitates understanding trade-offs between proprietary and open-source models, assessing both performance and deployment practicality. 

\section{Experimental Results}

We conducted experimental evaluations using the MovieLens 100K and 1M datasets, employing a 5-fold cross-validation framework with an 80:20 train-test split. Due to dataset sparsity (\~97\%), evaluations focused strictly on explicit user-item interactions to accurately measure model effectiveness. All code and resources are publicly accessible via our GitHub repository.\footnote{\url{https://github.com/anonymous-public/End_to_End_Personalization}}

Our proposed methodology consistently outperformed established baselines (NGCF, LightGCN, ALS, and a GAT model without LLM enhancements) across Precision, Recall, NDCG, and MAP metrics, as shown in Figure 2. Key performance improvements resulted primarily from two factors: LLM-generated profiles providing semantically rich embeddings and a cosine similarity term that enhanced latent space alignment. Moreover, structured unified textual representations consistently outperformed integrated textual representations, confirming structured schemas' efficacy in improving semantic coherence and embedding initialization.

Due to operational constraints, comprehensive post-processing evaluations using LLM architectures (Gemma-3 4B, OpenAI 4.1-mini, and o4-mini) were conducted only on the MovieLens 100K dataset. Surprisingly, OpenAI's 4.1-mini consistently achieved the best performance, surpassing the reasoning-focused o4-mini, which performed comparably to the Gemma-3 model despite its larger size. This result underscores Gemma-3's efficiency given its smaller parameter count.

Among the LLM-based post-processing strategies, the relevancy scoring method consistently yielded the highest Precision, NDCG, and MAP in the cold start scenario (users with less than ten interactions). Its strong performance can be attributed to averaging scores across multiple evaluations, reducing biases and enhancing stability. The pairwise Binary Search Tree (BST) approach ranked second, excelling notably in Recall and NDCG, demonstrating the reliability of simplified binary judgments from LLMs, especially beneficial in cold-start scenarios.

\begin{figure*}[t]
  \centering
  \includegraphics[width=0.75\textwidth, height=0.31\textheight, keepaspectratio=true]{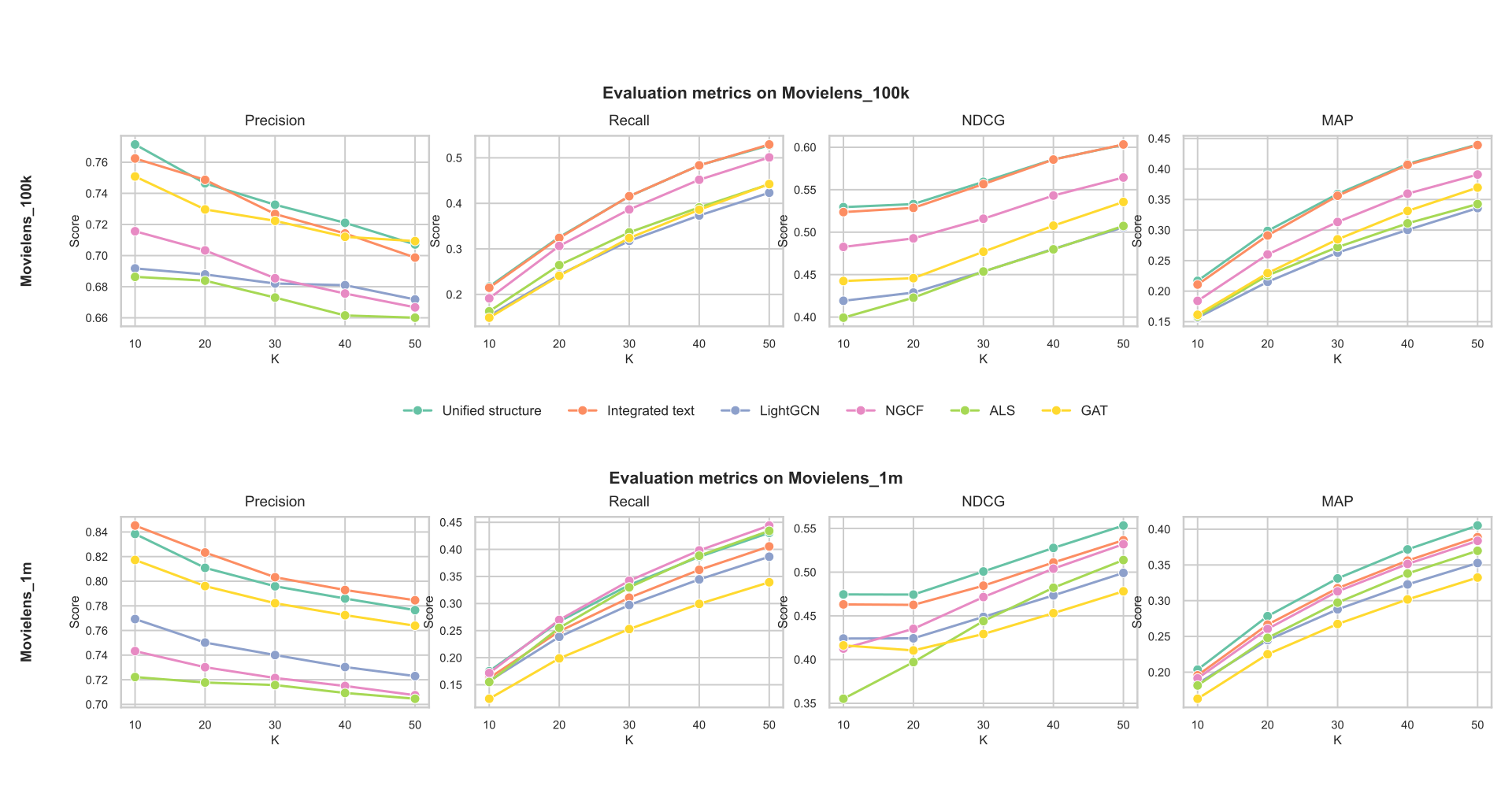}
  \caption{\small{Evaluation Metrics on Movielens 100k and 1M datasets.}}
  \Description{Movielens 100k and 1M evaluation plots comparing recommendation model performance across multiple metrics.}
  \label{fig:evaluation}
\end{figure*}

\begin{figure*}[t]
  \centering
  \includegraphics[width=0.75\textwidth, height=0.31\textheight, keepaspectratio=true]{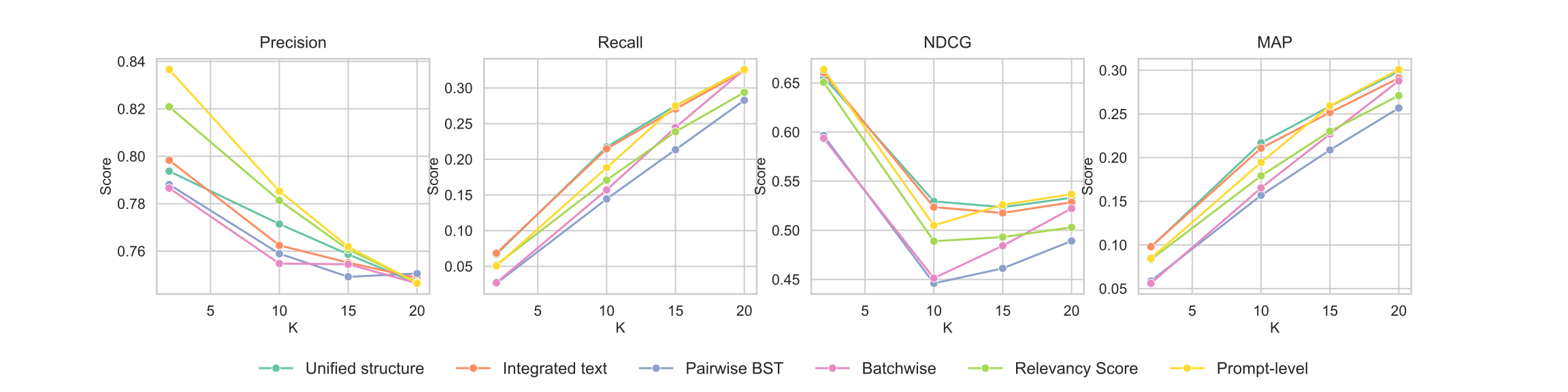}
 \caption{\footnotesize{Comparison between different post-processing methods on Movielens 100k.}}
  \label{fig:post_processing}
\end{figure*}

However, when excluding cold-start scenarios, our semantic re-ranking methods did not improve our baseline GAT method except for precision. This indicates that highly optimized embeddings from our core approach leave limited scope for further improvement through re-ranking when ample interaction data is present. Conversely, as shown in Table 1, semantic re-ranking substantially improved relevance and rankings in cold-start conditions, validating the critical role of semantic insights when user-item interaction data are sparse.

Additionally, we generated natural-language explanations for each recommendation using structured user and item profiles. Manual evaluation of selected recommendations confirmed these explanations as accurate, contextually relevant, and aligned with user preferences and historical interactions. However, the absence of standardized automated metrics highlights a need for further research into evaluation methodologies for explainable recommendations.

Key insights from our experiments include:
\begin{itemize}
    \item Structured profiles substantially improve semantic embedding quality.
    \item Relevancy scoring effectively balances precision and consistency.
    \item Pairwise BST comparisons offer strong recall and semantic reliability.
    \item Semantic re-ranking is especially beneficial in cold-start conditions, although less impactful with ample historical data.
    \item Standardized metrics for evaluating explanation quality remain a critical area for future research.
\end{itemize}

Future efforts should focus on balancing semantic depth, diversity, computational efficiency, fairness-aware constraints, and novel evaluation methods for explainability to enhance recommendation quality and user satisfaction.

\begin{table}[t]
  \centering
  \caption{Performance of cold-start users (fewer than 10 interactions) for Movielens100k and $k=10$.}
  \label{tab:cold_start_results}
  \begin{tabular}{lcccc}
    \toprule
    Method & Precision & Recall & NDCG & MAP \\
    \midrule
    \texttt{ALS} & 0.541 & 0.082 & 0.197 & 0.127 \\
    \texttt{NGCF} & 0.633 & 0.126 & 0.253 & 0.162 \\
    \texttt{LightGCN} & 0.700 & 0.121 & 0.347 & 0.196 \\
    \texttt{GAT } & 0.562 & 0.105 & 0.271 & 0.202 \\
    \texttt{Integrated text} & 0.555 & 0.241 & 0.314 & \textbf{0.319} \\
    \texttt{Unified structure} & 0.566 & 0.183 & 0.335 & 0.271 \\
    \texttt{Prompt-level} & 0.600 & 0.166 & 0.319 & 0.250 \\
    \texttt{Pairwise BST} & 0.693 & \textbf{0.244} & 0.380 & 0.234 \\
    \texttt{Batchwise} & 0.650 & 0.166 & 0.289 & 0.214 \\
    \texttt{Relevancy score} & \textbf{0.750} & 0.237 & \textbf{0.387} & 0.255  \\
    \bottomrule
  \end{tabular}
\end{table}

\section{Conclusion and Future Work}
Our study demonstrates the efficacy of systematically integrating LLMs into hybrid recommender systems, achieving state-of-the-art performance on MovieLens datasets. By synergizing LLM-generated semantic profiles with graph-based collaborative filtering, we address critical limitations in traditional systems, particularly in capturing nuanced user preferences and item characteristics. The iterative LLM-driven profile refinement, coupled with a GAT architecture optimized via BPR loss and cosine alignment, enables robust representation learning even under extreme sparsity. Post-processing with LLM-based re-ranking further enhances recommendation quality, outperforming baselines across precision, recall, and ranking metrics while maintaining explainability. Notably, our method excels in cold-start scenarios, proving its adaptability to sparse interaction data. The structured schema alignment and hybrid reranking strategies (e.g., BST-based comparisons) ensure semantic coherence while mitigating LLM hallucinations, validating the practicality of combining neural graph models with LLM reasoning.

The proposed method can be extended in various ways. One of these is mitigating biases introduced by metadata richness and enhancing diversity without sacrificing accuracy on more datasets. Techniques such as fairness-aware regularization, dynamic user preference adaptation, and lightweight LLM fine-tuning for domain-specific tasks could further optimize efficiency and scalability. Additionally, developing end-to-end training pipelines that jointly optimize graph embeddings and LLM reranking—rather than treating them as separate stages—could reduce computational overhead and improve alignment. Finally, user studies are needed to validate the perceived quality of explanations and ensure ethical transparency in LLM-driven recommendations.


\bibliographystyle{ACM-Reference-Format}

\begin{thebibliography}{38}


\ifx \showCODEN    \undefined \def \showCODEN     #1{\unskip}     \fi
\ifx \showISBNx    \undefined \def \showISBNx     #1{\unskip}     \fi
\ifx \showISBNxiii \undefined \def \showISBNxiii  #1{\unskip}     \fi
\ifx \showISSN     \undefined \def \showISSN      #1{\unskip}     \fi
\ifx \showLCCN     \undefined \def \showLCCN      #1{\unskip}     \fi
\ifx \shownote     \undefined \def \shownote      #1{#1}          \fi
\ifx \showarticletitle \undefined \def \showarticletitle #1{#1}   \fi
\ifx \showURL      \undefined \def \showURL       {\relax}        \fi
\providecommand\bibfield[2]{#2}
\providecommand\bibinfo[2]{#2}
\providecommand\natexlab[1]{#1}
\providecommand\showeprint[2][]{arXiv:#2}

\bibitem[Bao et~al\mbox{.}(2023)]%
        {DBLP:conf/recsys/BaoZZWF023}
\bibfield{author}{\bibinfo{person}{Keqin Bao}, \bibinfo{person}{Jizhi Zhang}, \bibinfo{person}{Yang Zhang}, \bibinfo{person}{Wenjie Wang}, \bibinfo{person}{Fuli Feng}, {and} \bibinfo{person}{Xiangnan He}.} \bibinfo{year}{2023}\natexlab{}.
\newblock \showarticletitle{TALLRec: An Effective and Efficient Tuning Framework to Align Large Language Model with Recommendation}. In \bibinfo{booktitle}{\emph{Proceedings of the 17th {ACM} Conference on Recommender Systems, RecSys 2023, Singapore, Singapore, September 18-22, 2023}}, \bibfield{editor}{\bibinfo{person}{Jie Zhang}, \bibinfo{person}{Li~Chen}, \bibinfo{person}{Shlomo Berkovsky}, \bibinfo{person}{Min Zhang}, \bibinfo{person}{Tommaso~Di Noia}, \bibinfo{person}{Justin Basilico}, \bibinfo{person}{Luiz Pizzato}, {and} \bibinfo{person}{Yang Song}} (Eds.). \bibinfo{publisher}{{ACM}}, \bibinfo{pages}{1007--1014}.
\newblock
\href{https://doi.org/10.1145/3604915.3608857}{doi:\nolinkurl{10.1145/3604915.3608857}}


\bibitem[Chen et~al\mbox{.}(2023)]%
        {DBLP:journals/corr/abs-2305-09858}
\bibfield{author}{\bibinfo{person}{Jiao Chen}, \bibinfo{person}{Luyi Ma}, \bibinfo{person}{Xiaohan Li}, \bibinfo{person}{Nikhil Thakurdesai}, \bibinfo{person}{Jianpeng Xu}, \bibinfo{person}{Jason H.~D. Cho}, \bibinfo{person}{Kaushiki Nag}, \bibinfo{person}{Evren K{\"{o}}rpeoglu}, \bibinfo{person}{Sushant Kumar}, {and} \bibinfo{person}{Kannan Achan}.} \bibinfo{year}{2023}\natexlab{}.
\newblock \showarticletitle{Knowledge Graph Completion Models are Few-shot Learners: An Empirical Study of Relation Labeling in E-commerce with LLMs}.
\newblock \bibinfo{journal}{\emph{CoRR}}  \bibinfo{volume}{abs/2305.09858} (\bibinfo{year}{2023}).
\newblock
\href{https://doi.org/10.48550/ARXIV.2305.09858}{doi:\nolinkurl{10.48550/ARXIV.2305.09858}}
\showeprint[arXiv]{2305.09858}


\bibitem[Cheng et~al\mbox{.}(2024)]%
        {DBLP:journals/fcsc/ChengLZLZC24}
\bibfield{author}{\bibinfo{person}{Mingyue Cheng}, \bibinfo{person}{Qi Liu}, \bibinfo{person}{Wenyu Zhang}, \bibinfo{person}{Zhiding Liu}, \bibinfo{person}{Hongke Zhao}, {and} \bibinfo{person}{Enhong Chen}.} \bibinfo{year}{2024}\natexlab{}.
\newblock \showarticletitle{A general tail item representation enhancement framework for sequential recommendation}.
\newblock \bibinfo{journal}{\emph{Frontiers Comput. Sci.}} \bibinfo{volume}{18}, \bibinfo{number}{6} (\bibinfo{year}{2024}), \bibinfo{pages}{186333}.
\newblock
\href{https://doi.org/10.1007/S11704-023-3112-Y}{doi:\nolinkurl{10.1007/S11704-023-3112-Y}}


\bibitem[Chu et~al\mbox{.}(2024)]%
        {DBLP:journals/corr/abs-2401-08217}
\bibfield{author}{\bibinfo{person}{Zhixuan Chu}, \bibinfo{person}{Yan Wang}, \bibinfo{person}{Qing Cui}, \bibinfo{person}{Longfei Li}, \bibinfo{person}{Wenqing Chen}, \bibinfo{person}{Sheng Li}, \bibinfo{person}{Zhan Qin}, {and} \bibinfo{person}{Kui Ren}.} \bibinfo{year}{2024}\natexlab{}.
\newblock \showarticletitle{LLM-Guided Multi-View Hypergraph Learning for Human-Centric Explainable Recommendation}.
\newblock \bibinfo{journal}{\emph{CoRR}}  \bibinfo{volume}{abs/2401.08217} (\bibinfo{year}{2024}).
\newblock
\href{https://doi.org/10.48550/ARXIV.2401.08217}{doi:\nolinkurl{10.48550/ARXIV.2401.08217}}
\showeprint[arXiv]{2401.08217}


\bibitem[Du et~al\mbox{.}(2024)]%
        {DBLP:conf/aaai/DuL0W0Z0Z24}
\bibfield{author}{\bibinfo{person}{Yingpeng Du}, \bibinfo{person}{Di Luo}, \bibinfo{person}{Rui Yan}, \bibinfo{person}{Xiaopei Wang}, \bibinfo{person}{Hongzhi Liu}, \bibinfo{person}{Hengshu Zhu}, \bibinfo{person}{Yang Song}, {and} \bibinfo{person}{Jie Zhang}.} \bibinfo{year}{2024}\natexlab{}.
\newblock \showarticletitle{Enhancing Job Recommendation through LLM-Based Generative Adversarial Networks}. In \bibinfo{booktitle}{\emph{Thirty-Eighth {AAAI} Conference on Artificial Intelligence, {AAAI} 2024, Thirty-Sixth Conference on Innovative Applications of Artificial Intelligence, {IAAI} 2024, Fourteenth Symposium on Educational Advances in Artificial Intelligence, {EAAI} 2014, February 20-27, 2024, Vancouver, Canada}}, \bibfield{editor}{\bibinfo{person}{Michael~J. Wooldridge}, \bibinfo{person}{Jennifer~G. Dy}, {and} \bibinfo{person}{Sriraam Natarajan}} (Eds.). \bibinfo{publisher}{{AAAI} Press}, \bibinfo{pages}{8363--8371}.
\newblock
\href{https://doi.org/10.1609/AAAI.V38I8.28678}{doi:\nolinkurl{10.1609/AAAI.V38I8.28678}}


\bibitem[Ebrat and Rueda(2024)]%
        {DBLP:journals/corr/abs-2405-13362}
\bibfield{author}{\bibinfo{person}{Danial Ebrat} {and} \bibinfo{person}{Luis Rueda}.} \bibinfo{year}{2024}\natexlab{}.
\newblock \showarticletitle{Lusifer: LLM-based User SImulated Feedback Environment for online Recommender systems}.
\newblock \bibinfo{journal}{\emph{CoRR}}  \bibinfo{volume}{abs/2405.13362} (\bibinfo{year}{2024}).
\newblock
\href{https://doi.org/10.48550/ARXIV.2405.13362}{doi:\nolinkurl{10.48550/ARXIV.2405.13362}}
\showeprint[arXiv]{2405.13362}


\bibitem[Elahi et~al\mbox{.}(2025)]%
        {DBLP:journals/eswa/ElahiAARNHW25}
\bibfield{author}{\bibinfo{person}{Ehsan Elahi}, \bibinfo{person}{Sajid Anwar}, \bibinfo{person}{Mousa Al{-}Kfairy}, \bibinfo{person}{Joel J. P.~C. Rodrigues}, \bibinfo{person}{Alladoumbaye Ngueilbaye}, \bibinfo{person}{Zahid Halim}, {and} \bibinfo{person}{Muhammad Waqas}.} \bibinfo{year}{2025}\natexlab{}.
\newblock \showarticletitle{Graph attention-based neural collaborative filtering for item-specific recommendation system using knowledge graph}.
\newblock \bibinfo{journal}{\emph{Expert Syst. Appl.}}  \bibinfo{volume}{266} (\bibinfo{year}{2025}), \bibinfo{pages}{126133}.
\newblock
\href{https://doi.org/10.1016/J.ESWA.2024.126133}{doi:\nolinkurl{10.1016/J.ESWA.2024.126133}}


\bibitem[Golub(1968)]%
        {golub1968least}
\bibfield{author}{\bibinfo{person}{Gene~Howard Golub}.} \bibinfo{year}{1968}\natexlab{}.
\newblock \showarticletitle{Least Squares, Singular Values and Matrix Approximations}.
\newblock \bibinfo{journal}{\emph{Aplikace matematiky}} \bibinfo{volume}{13}, \bibinfo{number}{1} (\bibinfo{year}{1968}), \bibinfo{pages}{44--51}.
\newblock
\showISSN{0373-6725}
\href{https://doi.org/10.21136/AM.1968.103138}{doi:\nolinkurl{10.21136/AM.1968.103138}}
\newblock
\shownote{\url{https://doi.org/10.21136/AM.1968.103138}}.


\bibitem[Gori and Pucci(2007)]%
        {DBLP:conf/ijcai/GoriP07}
\bibfield{author}{\bibinfo{person}{Marco Gori} {and} \bibinfo{person}{Augusto Pucci}.} \bibinfo{year}{2007}\natexlab{}.
\newblock \showarticletitle{ItemRank: {A} Random-Walk Based Scoring Algorithm for Recommender Engines}. In \bibinfo{booktitle}{\emph{{IJCAI} 2007, Proceedings of the 20th International Joint Conference on Artificial Intelligence, Hyderabad, India, January 6-12, 2007}}, \bibfield{editor}{\bibinfo{person}{Manuela~M. Veloso}} (Ed.). \bibinfo{pages}{2766--2771}.
\newblock
\urldef\tempurl%
\url{http://ijcai.org/Proceedings/07/Papers/444.pdf}
\showURL{%
\tempurl}


\bibitem[He et~al\mbox{.}(2020)]%
        {DBLP:conf/sigir/0001DWLZ020}
\bibfield{author}{\bibinfo{person}{Xiangnan He}, \bibinfo{person}{Kuan Deng}, \bibinfo{person}{Xiang Wang}, \bibinfo{person}{Yan Li}, \bibinfo{person}{Yong{-}Dong Zhang}, {and} \bibinfo{person}{Meng Wang}.} \bibinfo{year}{2020}\natexlab{}.
\newblock \showarticletitle{LightGCN: Simplifying and Powering Graph Convolution Network for Recommendation}. In \bibinfo{booktitle}{\emph{Proceedings of the 43rd International {ACM} {SIGIR} conference on research and development in Information Retrieval, {SIGIR} 2020, Virtual Event, China, July 25-30, 2020}}, \bibfield{editor}{\bibinfo{person}{Jimmy~X. Huang}, \bibinfo{person}{Yi~Chang}, \bibinfo{person}{Xueqi Cheng}, \bibinfo{person}{Jaap Kamps}, \bibinfo{person}{Vanessa Murdock}, \bibinfo{person}{Ji{-}Rong Wen}, {and} \bibinfo{person}{Yiqun Liu}} (Eds.). \bibinfo{publisher}{{ACM}}, \bibinfo{pages}{639--648}.
\newblock
\href{https://doi.org/10.1145/3397271.3401063}{doi:\nolinkurl{10.1145/3397271.3401063}}


\bibitem[He et~al\mbox{.}(2017a)]%
        {DBLP:journals/tkde/HeGKW17}
\bibfield{author}{\bibinfo{person}{Xiangnan He}, \bibinfo{person}{Ming Gao}, \bibinfo{person}{Min{-}Yen Kan}, {and} \bibinfo{person}{Dingxian Wang}.} \bibinfo{year}{2017}\natexlab{a}.
\newblock \showarticletitle{BiRank: Towards Ranking on Bipartite Graphs}.
\newblock \bibinfo{journal}{\emph{{IEEE} Trans. Knowl. Data Eng.}} \bibinfo{volume}{29}, \bibinfo{number}{1} (\bibinfo{year}{2017}), \bibinfo{pages}{57--71}.
\newblock
\href{https://doi.org/10.1109/TKDE.2016.2611584}{doi:\nolinkurl{10.1109/TKDE.2016.2611584}}


\bibitem[He et~al\mbox{.}(2017b)]%
        {DBLP:conf/www/HeLZNHC17}
\bibfield{author}{\bibinfo{person}{Xiangnan He}, \bibinfo{person}{Lizi Liao}, \bibinfo{person}{Hanwang Zhang}, \bibinfo{person}{Liqiang Nie}, \bibinfo{person}{Xia Hu}, {and} \bibinfo{person}{Tat{-}Seng Chua}.} \bibinfo{year}{2017}\natexlab{b}.
\newblock \showarticletitle{Neural Collaborative Filtering}. In \bibinfo{booktitle}{\emph{Proceedings of the 26th International Conference on World Wide Web, {WWW} 2017, Perth, Australia, April 3-7, 2017}}, \bibfield{editor}{\bibinfo{person}{Rick Barrett}, \bibinfo{person}{Rick Cummings}, \bibinfo{person}{Eugene Agichtein}, {and} \bibinfo{person}{Evgeniy Gabrilovich}} (Eds.). \bibinfo{publisher}{{ACM}}, \bibinfo{pages}{173--182}.
\newblock
\href{https://doi.org/10.1145/3038912.3052569}{doi:\nolinkurl{10.1145/3038912.3052569}}


\bibitem[Jiang et~al\mbox{.}(2024)]%
        {DBLP:conf/dasfaa/JiangQCLLZZZLLG24}
\bibfield{author}{\bibinfo{person}{Junzhe Jiang}, \bibinfo{person}{Shang Qu}, \bibinfo{person}{Mingyue Cheng}, \bibinfo{person}{Qi Liu}, \bibinfo{person}{Zhiding Liu}, \bibinfo{person}{Hao Zhang}, \bibinfo{person}{Rujiao Zhang}, \bibinfo{person}{Kai Zhang}, \bibinfo{person}{Rui Li}, \bibinfo{person}{Jiatong Li}, {and} \bibinfo{person}{Min Gao}.} \bibinfo{year}{2024}\natexlab{}.
\newblock \showarticletitle{Reformulating Sequential Recommendation: Learning Dynamic User Interest with Content-enriched Language Modeling}. In \bibinfo{booktitle}{\emph{Database Systems for Advanced Applications - 29th International Conference, {DASFAA} 2024, Gifu, Japan, July 2-5, 2024, Proceedings, Part {III}}} \emph{(\bibinfo{series}{Lecture Notes in Computer Science}, Vol.~\bibinfo{volume}{14852})}, \bibfield{editor}{\bibinfo{person}{Makoto Onizuka}, \bibinfo{person}{Jae{-}Gil Lee}, \bibinfo{person}{Yongxin Tong}, \bibinfo{person}{Chuan Xiao}, \bibinfo{person}{Yoshiharu Ishikawa}, \bibinfo{person}{Sihem Amer{-}Yahia}, \bibinfo{person}{H.~V. Jagadish}, {and} \bibinfo{person}{Kejing Lu}} (Eds.). \bibinfo{publisher}{Springer}, \bibinfo{pages}{353--362}.
\newblock
\href{https://doi.org/10.1007/978-981-97-5555-4\_25}{doi:\nolinkurl{10.1007/978-981-97-5555-4\_25}}


\bibitem[Lee and St{\"{o}}ger(2023)]%
        {DBLP:journals/simods/LeeS23}
\bibfield{author}{\bibinfo{person}{Kiryung Lee} {and} \bibinfo{person}{Dominik St{\"{o}}ger}.} \bibinfo{year}{2023}\natexlab{}.
\newblock \showarticletitle{Randomly Initialized Alternating Least Squares: Fast Convergence for Matrix Sensing}.
\newblock \bibinfo{journal}{\emph{{SIAM} J. Math. Data Sci.}} \bibinfo{volume}{5}, \bibinfo{number}{3} (\bibinfo{year}{2023}), \bibinfo{pages}{774--799}.
\newblock
\href{https://doi.org/10.1137/22M1506456}{doi:\nolinkurl{10.1137/22M1506456}}


\bibitem[Li et~al\mbox{.}(2023a)]%
        {DBLP:journals/corr/abs-2312-02443}
\bibfield{author}{\bibinfo{person}{Xinhang Li}, \bibinfo{person}{Chong Chen}, \bibinfo{person}{Xiangyu Zhao}, \bibinfo{person}{Yong Zhang}, {and} \bibinfo{person}{Chunxiao Xing}.} \bibinfo{year}{2023}\natexlab{a}.
\newblock \showarticletitle{E4SRec: An Elegant Effective Efficient Extensible Solution of Large Language Models for Sequential Recommendation}.
\newblock \bibinfo{journal}{\emph{CoRR}}  \bibinfo{volume}{abs/2312.02443} (\bibinfo{year}{2023}).
\newblock
\href{https://doi.org/10.48550/ARXIV.2312.02443}{doi:\nolinkurl{10.48550/ARXIV.2312.02443}}
\showeprint[arXiv]{2312.02443}


\bibitem[Li et~al\mbox{.}(2024)]%
        {DBLP:conf/aaai/LiMWHJ0X0J24}
\bibfield{author}{\bibinfo{person}{Yangning Li}, \bibinfo{person}{Shirong Ma}, \bibinfo{person}{Xiaobin Wang}, \bibinfo{person}{Shen Huang}, \bibinfo{person}{Chengyue Jiang}, \bibinfo{person}{Haitao Zheng}, \bibinfo{person}{Pengjun Xie}, \bibinfo{person}{Fei Huang}, {and} \bibinfo{person}{Yong Jiang}.} \bibinfo{year}{2024}\natexlab{}.
\newblock \showarticletitle{EcomGPT: Instruction-Tuning Large Language Models with Chain-of-Task Tasks for E-commerce}. In \bibinfo{booktitle}{\emph{Thirty-Eighth {AAAI} Conference on Artificial Intelligence, {AAAI} 2024, Thirty-Sixth Conference on Innovative Applications of Artificial Intelligence, {IAAI} 2024, Fourteenth Symposium on Educational Advances in Artificial Intelligence, {EAAI} 2014, February 20-27, 2024, Vancouver, Canada}}, \bibfield{editor}{\bibinfo{person}{Michael~J. Wooldridge}, \bibinfo{person}{Jennifer~G. Dy}, {and} \bibinfo{person}{Sriraam Natarajan}} (Eds.). \bibinfo{publisher}{{AAAI} Press}, \bibinfo{pages}{18582--18590}.
\newblock
\href{https://doi.org/10.1609/AAAI.V38I17.29820}{doi:\nolinkurl{10.1609/AAAI.V38I17.29820}}


\bibitem[Li et~al\mbox{.}(2023b)]%
        {DBLP:conf/emnlp/LiZL023}
\bibfield{author}{\bibinfo{person}{Zhuoyan Li}, \bibinfo{person}{Hangxiao Zhu}, \bibinfo{person}{Zhuoran Lu}, {and} \bibinfo{person}{Ming Yin}.} \bibinfo{year}{2023}\natexlab{b}.
\newblock \showarticletitle{Synthetic Data Generation with Large Language Models for Text Classification: Potential and Limitations}. In \bibinfo{booktitle}{\emph{Proceedings of the 2023 Conference on Empirical Methods in Natural Language Processing, {EMNLP} 2023, Singapore, December 6-10, 2023}}, \bibfield{editor}{\bibinfo{person}{Houda Bouamor}, \bibinfo{person}{Juan Pino}, {and} \bibinfo{person}{Kalika Bali}} (Eds.). \bibinfo{publisher}{Association for Computational Linguistics}, \bibinfo{pages}{10443--10461}.
\newblock
\href{https://doi.org/10.18653/V1/2023.EMNLP-MAIN.647}{doi:\nolinkurl{10.18653/V1/2023.EMNLP-MAIN.647}}


\bibitem[Lin et~al\mbox{.}(2024)]%
        {DBLP:conf/www/LinCWXQDZTY024}
\bibfield{author}{\bibinfo{person}{Jianghao Lin}, \bibinfo{person}{Bo Chen}, \bibinfo{person}{Hangyu Wang}, \bibinfo{person}{Yunjia Xi}, \bibinfo{person}{Yanru Qu}, \bibinfo{person}{Xinyi Dai}, \bibinfo{person}{Kangning Zhang}, \bibinfo{person}{Ruiming Tang}, \bibinfo{person}{Yong Yu}, {and} \bibinfo{person}{Weinan Zhang}.} \bibinfo{year}{2024}\natexlab{}.
\newblock \showarticletitle{ClickPrompt: {CTR} Models are Strong Prompt Generators for Adapting Language Models to {CTR} Prediction}. In \bibinfo{booktitle}{\emph{Proceedings of the {ACM} on Web Conference 2024, {WWW} 2024, Singapore, May 13-17, 2024}}, \bibfield{editor}{\bibinfo{person}{Tat{-}Seng Chua}, \bibinfo{person}{Chong{-}Wah Ngo}, \bibinfo{person}{Ravi Kumar}, \bibinfo{person}{Hady~W. Lauw}, {and} \bibinfo{person}{Roy~Ka{-}Wei Lee}} (Eds.). \bibinfo{publisher}{{ACM}}, \bibinfo{pages}{3319--3330}.
\newblock
\href{https://doi.org/10.1145/3589334.3645396}{doi:\nolinkurl{10.1145/3589334.3645396}}


\bibitem[Lin et~al\mbox{.}(2023)]%
        {DBLP:journals/corr/abs-2306-05817}
\bibfield{author}{\bibinfo{person}{Jianghao Lin}, \bibinfo{person}{Xinyi Dai}, \bibinfo{person}{Yunjia Xi}, \bibinfo{person}{Weiwen Liu}, \bibinfo{person}{Bo Chen}, \bibinfo{person}{Xiangyang Li}, \bibinfo{person}{Chenxu Zhu}, \bibinfo{person}{Huifeng Guo}, \bibinfo{person}{Yong Yu}, \bibinfo{person}{Ruiming Tang}, {and} \bibinfo{person}{Weinan Zhang}.} \bibinfo{year}{2023}\natexlab{}.
\newblock \showarticletitle{How Can Recommender Systems Benefit from Large Language Models: {A} Survey}.
\newblock \bibinfo{journal}{\emph{CoRR}}  \bibinfo{volume}{abs/2306.05817} (\bibinfo{year}{2023}).
\newblock
\href{https://doi.org/10.48550/ARXIV.2306.05817}{doi:\nolinkurl{10.48550/ARXIV.2306.05817}}
\showeprint[arXiv]{2306.05817}


\bibitem[Liu et~al\mbox{.}(2023b)]%
        {DBLP:journals/corr/abs-2312-16275}
\bibfield{author}{\bibinfo{person}{Fan Liu}, \bibinfo{person}{Yaqi Liu}, \bibinfo{person}{Zhiyong Cheng}, \bibinfo{person}{Liqiang Nie}, {and} \bibinfo{person}{Mohan~S. Kankanhalli}.} \bibinfo{year}{2023}\natexlab{b}.
\newblock \showarticletitle{Understanding Before Recommendation: Semantic Aspect-Aware Review Exploitation via Large Language Models}.
\newblock \bibinfo{journal}{\emph{CoRR}}  \bibinfo{volume}{abs/2312.16275} (\bibinfo{year}{2023}).
\newblock
\href{https://doi.org/10.48550/ARXIV.2312.16275}{doi:\nolinkurl{10.48550/ARXIV.2312.16275}}
\showeprint[arXiv]{2312.16275}


\bibitem[Liu et~al\mbox{.}(2023a)]%
        {DBLP:conf/recsys/Liu0LTW23}
\bibfield{author}{\bibinfo{person}{Weiwen Liu}, \bibinfo{person}{Wei Guo}, \bibinfo{person}{Yong Liu}, \bibinfo{person}{Ruiming Tang}, {and} \bibinfo{person}{Hao Wang}.} \bibinfo{year}{2023}\natexlab{a}.
\newblock \showarticletitle{User Behavior Modeling with Deep Learning for Recommendation: Recent Advances}. In \bibinfo{booktitle}{\emph{Proceedings of the 17th {ACM} Conference on Recommender Systems, RecSys 2023, Singapore, Singapore, September 18-22, 2023}}, \bibfield{editor}{\bibinfo{person}{Jie Zhang}, \bibinfo{person}{Li~Chen}, \bibinfo{person}{Shlomo Berkovsky}, \bibinfo{person}{Min Zhang}, \bibinfo{person}{Tommaso~Di Noia}, \bibinfo{person}{Justin Basilico}, \bibinfo{person}{Luiz Pizzato}, {and} \bibinfo{person}{Yang Song}} (Eds.). \bibinfo{publisher}{{ACM}}, \bibinfo{pages}{1286--1287}.
\newblock
\href{https://doi.org/10.1145/3604915.3609496}{doi:\nolinkurl{10.1145/3604915.3609496}}


\bibitem[Liu et~al\mbox{.}(2024)]%
        {DBLP:conf/www/LiuCZDWLL0024}
\bibfield{author}{\bibinfo{person}{Zhenghao Liu}, \bibinfo{person}{Zulong Chen}, \bibinfo{person}{Moufeng Zhang}, \bibinfo{person}{Shaoyang Duan}, \bibinfo{person}{Hong Wen}, \bibinfo{person}{Liangyue Li}, \bibinfo{person}{Nan Li}, \bibinfo{person}{Yu Gu}, {and} \bibinfo{person}{Ge Yu}.} \bibinfo{year}{2024}\natexlab{}.
\newblock \showarticletitle{Modeling User Viewing Flow using Large Language Models for Article Recommendation}. In \bibinfo{booktitle}{\emph{Companion Proceedings of the {ACM} on Web Conference 2024, {WWW} 2024, Singapore, Singapore, May 13-17, 2024}}, \bibfield{editor}{\bibinfo{person}{Tat{-}Seng Chua}, \bibinfo{person}{Chong{-}Wah Ngo}, \bibinfo{person}{Roy~Ka{-}Wei Lee}, \bibinfo{person}{Ravi Kumar}, {and} \bibinfo{person}{Hady~W. Lauw}} (Eds.). \bibinfo{publisher}{{ACM}}, \bibinfo{pages}{83--92}.
\newblock
\href{https://doi.org/10.1145/3589335.3648305}{doi:\nolinkurl{10.1145/3589335.3648305}}


\bibitem[Shi et~al\mbox{.}(2023)]%
        {DBLP:journals/corr/abs-2308-04913}
\bibfield{author}{\bibinfo{person}{Kaize Shi}, \bibinfo{person}{Xueyao Sun}, \bibinfo{person}{Dingxian Wang}, \bibinfo{person}{Yinlin Fu}, \bibinfo{person}{Guandong Xu}, {and} \bibinfo{person}{Qing Li}.} \bibinfo{year}{2023}\natexlab{}.
\newblock \showarticletitle{LLaMA-E: Empowering E-commerce Authoring with Multi-Aspect Instruction Following}.
\newblock \bibinfo{journal}{\emph{CoRR}}  \bibinfo{volume}{abs/2308.04913} (\bibinfo{year}{2023}).
\newblock
\href{https://doi.org/10.48550/ARXIV.2308.04913}{doi:\nolinkurl{10.48550/ARXIV.2308.04913}}
\showeprint[arXiv]{2308.04913}


\bibitem[Torbati et~al\mbox{.}(2023)]%
        {DBLP:journals/corr/abs-2311-01314}
\bibfield{author}{\bibinfo{person}{Ghazaleh~Haratinezhad Torbati}, \bibinfo{person}{Anna Tigunova}, \bibinfo{person}{Andrew Yates}, {and} \bibinfo{person}{Gerhard Weikum}.} \bibinfo{year}{2023}\natexlab{}.
\newblock \showarticletitle{Recommendations by Concise User Profiles from Review Text}.
\newblock \bibinfo{journal}{\emph{CoRR}}  \bibinfo{volume}{abs/2311.01314} (\bibinfo{year}{2023}).
\newblock
\href{https://doi.org/10.48550/ARXIV.2311.01314}{doi:\nolinkurl{10.48550/ARXIV.2311.01314}}
\showeprint[arXiv]{2311.01314}


\bibitem[Wang et~al\mbox{.}(2020)]%
        {DBLP:conf/nips/WangW0B0020}
\bibfield{author}{\bibinfo{person}{Wenhui Wang}, \bibinfo{person}{Furu Wei}, \bibinfo{person}{Li Dong}, \bibinfo{person}{Hangbo Bao}, \bibinfo{person}{Nan Yang}, {and} \bibinfo{person}{Ming Zhou}.} \bibinfo{year}{2020}\natexlab{}.
\newblock \showarticletitle{MiniLM: Deep Self-Attention Distillation for Task-Agnostic Compression of Pre-Trained Transformers}. In \bibinfo{booktitle}{\emph{Advances in Neural Information Processing Systems 33: Annual Conference on Neural Information Processing Systems 2020, NeurIPS 2020, December 6-12, 2020, virtual}}, \bibfield{editor}{\bibinfo{person}{Hugo Larochelle}, \bibinfo{person}{Marc'Aurelio Ranzato}, \bibinfo{person}{Raia Hadsell}, \bibinfo{person}{Maria{-}Florina Balcan}, {and} \bibinfo{person}{Hsuan{-}Tien Lin}} (Eds.).
\newblock
\urldef\tempurl%
\url{https://proceedings.neurips.cc/paper/2020/hash/3f5ee243547dee91fbd053c1c4a845aa-Abstract.html}
\showURL{%
\tempurl}


\bibitem[Wang et~al\mbox{.}(2019)]%
        {DBLP:conf/sigir/Wang0WFC19}
\bibfield{author}{\bibinfo{person}{Xiang Wang}, \bibinfo{person}{Xiangnan He}, \bibinfo{person}{Meng Wang}, \bibinfo{person}{Fuli Feng}, {and} \bibinfo{person}{Tat{-}Seng Chua}.} \bibinfo{year}{2019}\natexlab{}.
\newblock \showarticletitle{Neural Graph Collaborative Filtering}. In \bibinfo{booktitle}{\emph{Proceedings of the 42nd International {ACM} {SIGIR} Conference on Research and Development in Information Retrieval, {SIGIR} 2019, Paris, France, July 21-25, 2019}}, \bibfield{editor}{\bibinfo{person}{Benjamin Piwowarski}, \bibinfo{person}{Max Chevalier}, \bibinfo{person}{{\'{E}}ric Gaussier}, \bibinfo{person}{Yoelle Maarek}, \bibinfo{person}{Jian{-}Yun Nie}, {and} \bibinfo{person}{Falk Scholer}} (Eds.). \bibinfo{publisher}{{ACM}}, \bibinfo{pages}{165--174}.
\newblock
\href{https://doi.org/10.1145/3331184.3331267}{doi:\nolinkurl{10.1145/3331184.3331267}}


\bibitem[Wang et~al\mbox{.}(2023)]%
        {DBLP:journals/corr/abs-2308-10835}
\bibfield{author}{\bibinfo{person}{Yan Wang}, \bibinfo{person}{Zhixuan Chu}, \bibinfo{person}{Xin Ouyang}, \bibinfo{person}{Simeng Wang}, \bibinfo{person}{Hongyan Hao}, \bibinfo{person}{Yue Shen}, \bibinfo{person}{Jinjie Gu}, \bibinfo{person}{Siqiao Xue}, \bibinfo{person}{James~Y. Zhang}, \bibinfo{person}{Qing Cui}, \bibinfo{person}{Longfei Li}, \bibinfo{person}{Jun Zhou}, {and} \bibinfo{person}{Sheng Li}.} \bibinfo{year}{2023}\natexlab{}.
\newblock \showarticletitle{Enhancing Recommender Systems with Large Language Model Reasoning Graphs}.
\newblock \bibinfo{journal}{\emph{CoRR}}  \bibinfo{volume}{abs/2308.10835} (\bibinfo{year}{2023}).
\newblock
\href{https://doi.org/10.48550/ARXIV.2308.10835}{doi:\nolinkurl{10.48550/ARXIV.2308.10835}}
\showeprint[arXiv]{2308.10835}


\bibitem[Wei et~al\mbox{.}(2024)]%
        {DBLP:conf/wsdm/WeiRTWSCWYH24}
\bibfield{author}{\bibinfo{person}{Wei Wei}, \bibinfo{person}{Xubin Ren}, \bibinfo{person}{Jiabin Tang}, \bibinfo{person}{Qinyong Wang}, \bibinfo{person}{Lixin Su}, \bibinfo{person}{Suqi Cheng}, \bibinfo{person}{Junfeng Wang}, \bibinfo{person}{Dawei Yin}, {and} \bibinfo{person}{Chao Huang}.} \bibinfo{year}{2024}\natexlab{}.
\newblock \showarticletitle{LLMRec: Large Language Models with Graph Augmentation for Recommendation}. In \bibinfo{booktitle}{\emph{Proceedings of the 17th {ACM} International Conference on Web Search and Data Mining, {WSDM} 2024, Merida, Mexico, March 4-8, 2024}}, \bibfield{editor}{\bibinfo{person}{Luz~Angelica Caudillo{-}Mata}, \bibinfo{person}{Silvio Lattanzi}, \bibinfo{person}{Andr{\'{e}}s~Mu{\~{n}}oz Medina}, \bibinfo{person}{Leman Akoglu}, \bibinfo{person}{Aristides Gionis}, {and} \bibinfo{person}{Sergei Vassilvitskii}} (Eds.). \bibinfo{publisher}{{ACM}}, \bibinfo{pages}{806--815}.
\newblock
\href{https://doi.org/10.1145/3616855.3635853}{doi:\nolinkurl{10.1145/3616855.3635853}}


\bibitem[Wu et~al\mbox{.}(2024)]%
        {DBLP:journals/www/WuZQWGSQZZLXC24}
\bibfield{author}{\bibinfo{person}{Likang Wu}, \bibinfo{person}{Zhi Zheng}, \bibinfo{person}{Zhaopeng Qiu}, \bibinfo{person}{Hao Wang}, \bibinfo{person}{Hongchao Gu}, \bibinfo{person}{Tingjia Shen}, \bibinfo{person}{Chuan Qin}, \bibinfo{person}{Chen Zhu}, \bibinfo{person}{Hengshu Zhu}, \bibinfo{person}{Qi Liu}, \bibinfo{person}{Hui Xiong}, {and} \bibinfo{person}{Enhong Chen}.} \bibinfo{year}{2024}\natexlab{}.
\newblock \showarticletitle{A survey on large language models for recommendation}.
\newblock \bibinfo{journal}{\emph{World Wide Web {(WWW)}}} \bibinfo{volume}{27}, \bibinfo{number}{5} (\bibinfo{year}{2024}), \bibinfo{pages}{60}.
\newblock
\href{https://doi.org/10.1007/S11280-024-01291-2}{doi:\nolinkurl{10.1007/S11280-024-01291-2}}


\bibitem[Xi et~al\mbox{.}(2024)]%
        {DBLP:conf/recsys/XiLLCZZCT0024}
\bibfield{author}{\bibinfo{person}{Yunjia Xi}, \bibinfo{person}{Weiwen Liu}, \bibinfo{person}{Jianghao Lin}, \bibinfo{person}{Xiaoling Cai}, \bibinfo{person}{Hong Zhu}, \bibinfo{person}{Jieming Zhu}, \bibinfo{person}{Bo Chen}, \bibinfo{person}{Ruiming Tang}, \bibinfo{person}{Weinan Zhang}, {and} \bibinfo{person}{Yong Yu}.} \bibinfo{year}{2024}\natexlab{}.
\newblock \showarticletitle{Towards Open-World Recommendation with Knowledge Augmentation from Large Language Models}. In \bibinfo{booktitle}{\emph{Proceedings of the 18th {ACM} Conference on Recommender Systems, RecSys 2024, Bari, Italy, October 14-18, 2024}}, \bibfield{editor}{\bibinfo{person}{Tommaso~Di Noia}, \bibinfo{person}{Pasquale Lops}, \bibinfo{person}{Thorsten Joachims}, \bibinfo{person}{Katrien Verbert}, \bibinfo{person}{Pablo Castells}, \bibinfo{person}{Zhenhua Dong}, {and} \bibinfo{person}{Ben London}} (Eds.). \bibinfo{publisher}{{ACM}}, \bibinfo{pages}{12--22}.
\newblock
\href{https://doi.org/10.1145/3640457.3688104}{doi:\nolinkurl{10.1145/3640457.3688104}}


\bibitem[Yang et~al\mbox{.}(2023b)]%
        {DBLP:conf/bibm/YangZDXDLSJZL23}
\bibfield{author}{\bibinfo{person}{Qiang Yang}, \bibinfo{person}{Shuxin Zhang}, \bibinfo{person}{Suyu Dong}, \bibinfo{person}{Long Xu}, \bibinfo{person}{Weihe Dong}, \bibinfo{person}{Xiaokun Li}, \bibinfo{person}{Pengzhong Sun}, \bibinfo{person}{Feng Jiang}, \bibinfo{person}{Xianyu Zhang}, {and} \bibinfo{person}{Gongning Luo}.} \bibinfo{year}{2023}\natexlab{b}.
\newblock \showarticletitle{Graph Convolutional Network with Neural Inductive Matrix Completion for Predicting Disease-Related LncRNA Genes}. In \bibinfo{booktitle}{\emph{{IEEE} International Conference on Bioinformatics and Biomedicine, {BIBM} 2023, Istanbul, Turkiye, December 5-8, 2023}}, \bibfield{editor}{\bibinfo{person}{Xingpeng Jiang}, \bibinfo{person}{Haiying Wang}, \bibinfo{person}{Reda Alhajj}, \bibinfo{person}{Xiaohua Hu}, \bibinfo{person}{Felix Engel}, \bibinfo{person}{Mufti Mahmud}, \bibinfo{person}{Nadia Pisanti}, \bibinfo{person}{Xuefeng Cui}, {and} \bibinfo{person}{Hong Song}} (Eds.). \bibinfo{publisher}{{IEEE}}, \bibinfo{pages}{3595--3601}.
\newblock
\href{https://doi.org/10.1109/BIBM58861.2023.10386047}{doi:\nolinkurl{10.1109/BIBM58861.2023.10386047}}


\bibitem[Yang et~al\mbox{.}(2023a)]%
        {DBLP:conf/icdm/00040LXM0ZZFD23}
\bibfield{author}{\bibinfo{person}{Shenghao Yang}, \bibinfo{person}{Chenyang Wang}, \bibinfo{person}{Yankai Liu}, \bibinfo{person}{Kangping Xu}, \bibinfo{person}{Weizhi Ma}, \bibinfo{person}{Yiqun Liu}, \bibinfo{person}{Min Zhang}, \bibinfo{person}{Haitao Zeng}, \bibinfo{person}{Junlan Feng}, {and} \bibinfo{person}{Chao Deng}.} \bibinfo{year}{2023}\natexlab{a}.
\newblock \showarticletitle{Collaborative Word-based Pre-trained Item Representation for Transferable Recommendation}. In \bibinfo{booktitle}{\emph{{IEEE} International Conference on Data Mining, {ICDM} 2023, Shanghai, China, December 1-4, 2023}}, \bibfield{editor}{\bibinfo{person}{Guihai Chen}, \bibinfo{person}{Latifur Khan}, \bibinfo{person}{Xiaofeng Gao}, \bibinfo{person}{Meikang Qiu}, \bibinfo{person}{Witold Pedrycz}, {and} \bibinfo{person}{Xindong Wu}} (Eds.). \bibinfo{publisher}{{IEEE}}, \bibinfo{pages}{728--737}.
\newblock
\href{https://doi.org/10.1109/ICDM58522.2023.00082}{doi:\nolinkurl{10.1109/ICDM58522.2023.00082}}


\bibitem[Ying et~al\mbox{.}(2018)]%
        {DBLP:conf/kdd/YingHCEHL18}
\bibfield{author}{\bibinfo{person}{Rex Ying}, \bibinfo{person}{Ruining He}, \bibinfo{person}{Kaifeng Chen}, \bibinfo{person}{Pong Eksombatchai}, \bibinfo{person}{William~L. Hamilton}, {and} \bibinfo{person}{Jure Leskovec}.} \bibinfo{year}{2018}\natexlab{}.
\newblock \showarticletitle{Graph Convolutional Neural Networks for Web-Scale Recommender Systems}. In \bibinfo{booktitle}{\emph{Proceedings of the 24th {ACM} {SIGKDD} International Conference on Knowledge Discovery {\&} Data Mining, {KDD} 2018, London, UK, August 19-23, 2018}}, \bibfield{editor}{\bibinfo{person}{Yike Guo} {and} \bibinfo{person}{Faisal Farooq}} (Eds.). \bibinfo{publisher}{{ACM}}, \bibinfo{pages}{974--983}.
\newblock
\href{https://doi.org/10.1145/3219819.3219890}{doi:\nolinkurl{10.1145/3219819.3219890}}


\bibitem[Yue et~al\mbox{.}(2023)]%
        {DBLP:journals/corr/abs-2311-02089}
\bibfield{author}{\bibinfo{person}{Zhenrui Yue}, \bibinfo{person}{Sara Rabhi}, \bibinfo{person}{Gabriel de Souza Pereira~Moreira}, \bibinfo{person}{Dong Wang}, {and} \bibinfo{person}{Even Oldridge}.} \bibinfo{year}{2023}\natexlab{}.
\newblock \showarticletitle{LlamaRec: Two-Stage Recommendation using Large Language Models for Ranking}.
\newblock \bibinfo{journal}{\emph{CoRR}}  \bibinfo{volume}{abs/2311.02089} (\bibinfo{year}{2023}).
\newblock
\href{https://doi.org/10.48550/ARXIV.2311.02089}{doi:\nolinkurl{10.48550/ARXIV.2311.02089}}
\showeprint[arXiv]{2311.02089}


\bibitem[Zhang et~al\mbox{.}(2023)]%
        {DBLP:conf/adma/ZhangLWCG23}
\bibfield{author}{\bibinfo{person}{Shaowei Zhang}, \bibinfo{person}{Zhao Li}, \bibinfo{person}{Xin Wang}, \bibinfo{person}{Zirui Chen}, {and} \bibinfo{person}{Wenbin Guo}.} \bibinfo{year}{2023}\natexlab{}.
\newblock \showarticletitle{{TKGAT:} Temporal Knowledge Graph Representation Learning Using Attention Network}. In \bibinfo{booktitle}{\emph{Advanced Data Mining and Applications - 19th International Conference, {ADMA} 2023, Shenyang, China, August 21-23, 2023, Proceedings, Part {II}}} \emph{(\bibinfo{series}{Lecture Notes in Computer Science}, Vol.~\bibinfo{volume}{14177})}, \bibfield{editor}{\bibinfo{person}{Xiaochun Yang}, \bibinfo{person}{Heru Suhartanto}, \bibinfo{person}{Guoren Wang}, \bibinfo{person}{Bin Wang}, \bibinfo{person}{Jing Jiang}, \bibinfo{person}{Bing Li}, \bibinfo{person}{Huaijie Zhu}, {and} \bibinfo{person}{Ningning Cui}} (Eds.). \bibinfo{publisher}{Springer}, \bibinfo{pages}{46--61}.
\newblock
\href{https://doi.org/10.1007/978-3-031-46664-9\_4}{doi:\nolinkurl{10.1007/978-3-031-46664-9\_4}}


\bibitem[Zhao et~al\mbox{.}(2024)]%
        {DBLP:journals/tkde/ZhaoFLLMWWWZTL24}
\bibfield{author}{\bibinfo{person}{Zihuai Zhao}, \bibinfo{person}{Wenqi Fan}, \bibinfo{person}{Jiatong Li}, \bibinfo{person}{Yunqing Liu}, \bibinfo{person}{Xiaowei Mei}, \bibinfo{person}{Yiqi Wang}, \bibinfo{person}{Zhen Wen}, \bibinfo{person}{Fei Wang}, \bibinfo{person}{Xiangyu Zhao}, \bibinfo{person}{Jiliang Tang}, {and} \bibinfo{person}{Qing Li}.} \bibinfo{year}{2024}\natexlab{}.
\newblock \showarticletitle{Recommender Systems in the Era of Large Language Models (LLMs)}.
\newblock \bibinfo{journal}{\emph{{IEEE} Trans. Knowl. Data Eng.}} \bibinfo{volume}{36}, \bibinfo{number}{11} (\bibinfo{year}{2024}), \bibinfo{pages}{6889--6907}.
\newblock
\href{https://doi.org/10.1109/TKDE.2024.3392335}{doi:\nolinkurl{10.1109/TKDE.2024.3392335}}


\bibitem[Zheng et~al\mbox{.}(2018)]%
        {DBLP:conf/recsys/ZhengLJZY18}
\bibfield{author}{\bibinfo{person}{Lei Zheng}, \bibinfo{person}{Chun{-}Ta Lu}, \bibinfo{person}{Fei Jiang}, \bibinfo{person}{Jiawei Zhang}, {and} \bibinfo{person}{Philip~S. Yu}.} \bibinfo{year}{2018}\natexlab{}.
\newblock \showarticletitle{Spectral collaborative filtering}. In \bibinfo{booktitle}{\emph{Proceedings of the 12th {ACM} Conference on Recommender Systems, RecSys 2018, Vancouver, BC, Canada, October 2-7, 2018}}, \bibfield{editor}{\bibinfo{person}{Sole Pera}, \bibinfo{person}{Michael~D. Ekstrand}, \bibinfo{person}{Xavier Amatriain}, {and} \bibinfo{person}{John O'Donovan}} (Eds.). \bibinfo{publisher}{{ACM}}, \bibinfo{pages}{311--319}.
\newblock
\href{https://doi.org/10.1145/3240323.3240343}{doi:\nolinkurl{10.1145/3240323.3240343}}


\bibitem[Zheng et~al\mbox{.}(2023)]%
        {DBLP:journals/corr/abs-2307-02157}
\bibfield{author}{\bibinfo{person}{Zhi Zheng}, \bibinfo{person}{Zhaopeng Qiu}, \bibinfo{person}{Xiao Hu}, \bibinfo{person}{Likang Wu}, \bibinfo{person}{Hengshu Zhu}, {and} \bibinfo{person}{Hui Xiong}.} \bibinfo{year}{2023}\natexlab{}.
\newblock \showarticletitle{Generative Job Recommendations with Large Language Model}.
\newblock \bibinfo{journal}{\emph{CoRR}}  \bibinfo{volume}{abs/2307.02157} (\bibinfo{year}{2023}).
\newblock
\href{https://doi.org/10.48550/ARXIV.2307.02157}{doi:\nolinkurl{10.48550/ARXIV.2307.02157}}
\showeprint[arXiv]{2307.02157}


\end{thebibliography}


\end{document}